\documentclass[pra, 12pt, eqsecnum]{revtex4}
\usepackage[dvips]{graphics}

\newcommand {\psig}{\Sigma_}
\newcommand {\pxi}{\Xi_}
\newcommand {\apsig}[1]{\langle \Sigma_{#1} \rangle}
\newcommand {\apxi}[1]{\langle \Xi_{#1} \rangle}
\newcommand {\apsigxi}[2]{\langle \Sigma_{#1} \Xi_{#2} \rangle}
\newcommand {\avecsig}{\langle \vec{\Sigma} \rangle}
\begin{document}
\title{Dynamics of initially entangled open quantum systems}

\author{Thomas F. Jordan}
\email[email: ]{tjordan@d.umn.edu}
\affiliation{Physics Department, University of Minnesota, Duluth, Minnesota 55812}
\author{Anil Shaji}
\email[email: ]{shaji@physics.utexas.edu}
\affiliation{The University of Texas at Austin, Center for Statistical Mechanics, 1 University Station C1609, Austin TX 78712}
\author{E. C. G. Sudarshan}
\email[email: ]{sudarshan@physics.utexas.edu}
\affiliation{The University of Texas at Austin, Center for Particle Physics, 1 University Station C1602, Austin TX 78712}  

\begin{abstract}
Linear maps of matrices describing evolution of density matrices for a quantum system initially entangled with another are identified and found to be not always completely positive. They can even map a positive matrix to a matrix that is not positive, unless we restrict the domain on which the map acts. Nevertheless, their form is similar to that of completely positive maps. Only some minus signs are inserted in the operator-sum representation. Each map is the difference of two completely positive maps. The maps are first obtained as maps of mean values and then as maps of basis matrices. These forms also prove to be useful. An example for two entangled qubits is worked out in detail. Relation to earlier work is discussed.
\end{abstract}


\maketitle

\section{Introduction}

Linear maps of matrices can describe evolution of density matrices for a quantum system that interacts and entangles with another system \cite{sudarshan61a,jordan61,jordan62}. The simplest case is when the density matrix for the initial state of the combined system is a product of density matrices for the individual systems. Then the evolution of the single system can be described by a completely positive map. These maps have been extensively studied and used \cite{stormer63,choi72,choi74,davies76,kraus83,breuer02}. Here we consider the general case where the two systems may be entangled in the initial state. We ask what kind of map, if any, can describe the physics then. Completely positive maps can be used in quantum information processing because, with ability to decohere a system from its surroundings and initialize particular states, the two systems can be made separate so they have not been interacting and are not entangled when they are brought together in the initial state. What happens, though, when they are already entangled at the start?

We find that evolution can generally be described by linear maps of matrices. They are not completely positive maps. They can even map a positive matrix to a matrix that is not positive. Nevertheless, basic forms of the maps are similar to those of completely positive maps. Only some minus signs are inserted in the operator-sum representation. Each map is the difference of two completely positive maps. These familiar forms follow simply from the fact that the map takes every Hermitian matrix to a Hermitian matrix. The maps are first obtained as maps of mean values and then as maps of basis matrices. These forms also prove to be useful.

A new feature is that each map is made to be used for a particular set of states, to act in a particular domain. This is the set of states of the single system described by varying mean values of quantities for that system that are compatible with fixed mean values of other quantities for the combined system in describing an initial state of the combined system. We call that the {\em compatibility domain}. The map is defined for all matrices for the single system. In a domain that is larger than the compatibility domain, but still limited, every positive matrix is mapped to a positive matrix. We call that the {\em positivity domain}. We describe both domains for our example.

To extract the map that describes the evolution of one system from the dynamics of the two combined systems, we calculate changes of mean values (expectation values) in the Heisenberg picture. This allows us to hold calculations to the minimum needed to find the changes in the quantities that describe the single system. To make clear what we are doing, we keep our focus on those quantities and keep them separate from the other quantities in the description of the combined system, which may be parameters in the map.

There has been recognition of the limitations of completely positive maps in describing evolution of open quantum systems \cite{chuang97}, but little effort has been made to use more general maps there. Other considerations, including descriptions of entanglement and separability, have motivated substantial mathematical work on maps that are not completely positive but do take every positive matrix to a positive matrix \cite{jamiolkowski72,terhal01,arrighi04}. The maps we consider here do not need to have even that property.

We begin with an example for two entangled qubits, which we work out in detail. Then we outline the extension to any system described by finite matrices. In the concluding section we discuss how what is done here relates to earlier work \cite{pechukas94,alicki95,pechukas95} and point out the errors in arguments that a map describing evolution of an open quantum system has to be completely positive.

\section{Two-qubit examples}\label{sec2}

Consider two qubits described by two sets of Pauli matrices $\psig1$, $\psig2$, $\psig3$ and $\pxi1$, $\pxi2$, $\pxi3$. Let the Hamiltonian be
\begin{equation}
  \label{eq:hamil1}
  H = \frac{1}{2} \omega \psig3 \pxi1 .
\end{equation}
The evolution of the $\Sigma$ qubit is described by the mean values $\apsig{1}$, $\apsig{2}$ and $\apsig{3}$ at time zero changing to
\begin{eqnarray}
  \label{eq:evol1}
  \langle e^{iHt} \psig1 e^{-iHt} \rangle & = & \apsig{1} \cos \omega t - \apsigxi{2}{1} \sin \omega t \nonumber \\
  \langle e^{iHt} \psig2 e^{-iHt} \rangle & = & \apsig{2} \cos \omega t + \apsigxi{1}{1} \sin \omega t \nonumber \\
\langle e^{iHt} \psig3 e^{-iHt} \rangle & = & \apsig{3}
\end{eqnarray}
at time $t$. These three mean values describe the state of the $\Sigma$ qubit at time $t$.

\subsection{The basics for one time}\label{sec2a}

Look at this when $\omega t$ is $\pi /2$. Then the mean values are changed to 
\begin{equation}
  \label{eq:mean1}
  \apsig{1}' = a_1 \; , \qquad \apsig{2}' =a_2 \; , \qquad \apsig{3}' = \apsig{3}
\end{equation}
where
\begin{equation}
  \label{eq:param1}
  a_1 = -\apsigxi{2}{1} \; , \qquad a_2 = \apsigxi{1}{1}. 
\end{equation}
We consider the $a_1$, $a_2$ to be parameters that describe the effect of the dynamics of the two qubits that drives the evolution of the $\Sigma$ qubit, not quantities that are part of the description of the initial state of the $\Sigma$ qubit. What we do will apply to different initial states of the $\Sigma$ qubit for the same fixed $a_1$, $a_2$.

The change of mean values calculated in the Heisenberg picture determines the change of the density matrix in the Schr\"{o}dinger picture. The density matrix
\begin{equation}
  \label{eq:density1}
  \rho = \frac{1}{2} ( 1 + \langle \vec{\Sigma} \rangle \cdot \vec{\Sigma} )
\end{equation}
that describes the state of the $\Sigma$ qubit at time zero is changed to the density matrix
\begin{equation}
  \label{eq:density2}
  \rho' =  \frac{1}{2} ( 1 + \langle \vec{\Sigma} \rangle' \cdot \vec{\Sigma} )= \frac{1}{2} ( 1 + a_1 \psig1 + a_2 \psig2 + \apsig{3}\psig3)
\end{equation}
that describes the state of the $\Sigma$ qubit when $\omega t$ is $\pi/2$. This is the same for all the different $\avecsig$ that are compatible with the same fixed $\apsigxi{2}{1}$ and $\apsigxi{1}{1}$ in describing a possible initial state for the two qubits. We will refer to these as the {\em compatible} $\avecsig$. 

To be meaningful, a map has to act on a {\em substantial set} of states. To insure that we have something substantial to consider here, we will assume that the set of compatible $\avecsig$ is substantial. We will exclude those values of $\apsigxi{2}{1}$ and $\apsigxi{1}{1}$ that do not at least allow three-dimensional variation in the directions of the compatible $\avecsig$. For example, we will not let $\apsigxi{1}{1}$ be $1$, because that would imply $\apsig{2}$ and $\apsig{3}$ are zero. The set of compatible $\avecsig$ will be described more completely in Section \ref{domains}.

The change of density matrices can be extended to a linear map of all $2 \times 2$ matrices to $2 \times 2$ matrices defined by 
\begin{equation}
  \label{eq:map1}
  1' = 1+ a_1 \psig1 + a_2 \psig2 \; , \qquad \psig1' =0 \; , \qquad \psig2'=0 \; , \qquad \psig3'=\psig3\;.
\end{equation}
This takes each density matrix $\rho$ described by equation (\ref{eq:density1}), for each compatible $\avecsig$ in each different direction, to the density matrix
\begin{equation}
  \label{eq:density3}
  \rho' = \frac{1}{2} (1' + \avecsig \cdot \vec{\Sigma}')
\end{equation}
that is the same as that described by equation (\ref{eq:density2}). This map takes every Hermitian matrix to a Hermitian matrix. It does not map every positive matrix to a positive matrix. 

The map takes 
\begin{equation}
  \label{eq:op1}
  P = \frac{1}{2}(1 + \psig3)
\end{equation}
which is positive, to 
\begin{equation}
  \label{eq:op2}
  P' = \frac{1}{2}(1 + a_1 \psig1 + a_2 \psig2 + \psig3 )
\end{equation}
which is not positive. To see that $P'$ is not positive, let 
\begin{equation}
  \label{eq:rand1}
  a_1 = r \cos \theta \; , \qquad a_2 = r \sin \theta
\end{equation}
choose a vector $\psi$ such that
\begin{eqnarray}
  \label{eq:rand2}
  (\psi \, , \; 1 \, \psi) = 1 \; ,& \qquad  ( \psi \, , \; \psig1 \, \psi) =  -r \cos \theta/\sqrt{1 + r^2} \nonumber \\
 ( \psi \, , \; \psig2 \, \psi) = -r \sin \theta/\sqrt{1 + r^2}\; , &  \qquad ( \psi \, , \; \psig3 \, \psi) = -1/\sqrt{1 + r^2} 
\end{eqnarray}
and calculate 
\begin{equation}
  \label{eq:rand3}
  (\psi \, , \; P' \, \psi) = \frac{1}{2}( 1 - \sqrt{1 + r^2} ).
\end{equation}
This is negative even when $r$ is very small so that $\apsigxi{1}{2}$ and $\apsigxi{1}{1}$ are very small and there is room for a large set of compatible $\avecsig$.

Of course if $\rho$ is a density matrix that gives a compatible mean value $\avecsig$, the map takes $\rho$, described by equation (\ref{eq:density1}), to the density matrix $\rho'$ described by equation (\ref{eq:density2}), which is positive. To see explicitly that $\rho'$ is positive, consider that for any vector $\psi$ 
\begin{equation}
  \label{eq:pos1}
  |(\psi \, , \; \psig1 \, \psi)|^2 +  |(\psi \, , \; \psig2 \, \psi)|^2 +   |(\psi \, , \; \psig3 \, \psi)|^2    \leq |(\psi \, , \; \psi)|^2 
\end{equation}
and if $\apsig{3}$ is compatible with  $\apsigxi{1}{2}$ and $\apsigxi{1}{1}$ in describing a possible state for the two qubits, then 
\begin{equation}
  \label{eq:pos2}
  (a_1)^2 + (a_2)^2 + \apsig{3}^2 = \apsigxi{2}{1}^2 + \apsigxi{1}{1}^2 + \apsig{3}^2 \leq 1
\end{equation}
so that altogether
\begin{equation}
  \label{eq:pos3}
  |a_1(\psi \, , \; \psig1 \, \psi)+a_2(\psi \, , \; \psig2 \, \psi)+ \apsig{3}(\psi \, , \; \psig3 \, \psi)| \leq ( \psi \, , \; 1 \, \psi).
\end{equation}
The important difference between the density matrix $\rho$ and the positive matrix $P$ described by equation (\ref{eq:op1}) is the factor $\apsig{3}$ multiplying $\Sigma_3$ in the density matrix. If $\apsig{3}$ is changed to $1$, the inequality (\ref{eq:pos2}) can fail. The map can fail to take positive matrices to positive matrices when it extends beyond density matrices for compatible $\avecsig$. 

The map can fail to be completely positive even within the limits of compatible $\avecsig$ where it maps every positive matrix to a positive matrix. To see that, we extend the map to the two qubits by taking its product with the identity map of the matrices $1$, $\pxi1$, $\pxi2$, $\pxi3$. We have used equations (\ref{eq:map1}) to describe a map of $2 \times 2$ matrices. Now we use it to describe a map of $4 \times 4$ matrices; each matrix in equations (\ref{eq:map1}) is the product of the $2 \times 2$ matrix for the $\Sigma$ qubit with the identity matrix for the $\Xi$ qubit. In addition we get 
\begin{eqnarray}
  \label{eq:ncp1}
  (\Sigma_1 \Xi_k)'& = \psig1' \pxi k' &= 0, \nonumber \\
  (\Sigma_2 \Xi_k)'& = \psig2' \pxi k' &= 0, \nonumber \\
  (\Sigma_3 \Xi_k)'& = \psig3' \pxi k' &= \psig3 \pxi k, \nonumber \\
  \pxi k'=(1 \cdot \Xi_k)'& = 1' \pxi k' &= (1+a_1 \psig1+a_2 \psig2)\pxi k,
\end{eqnarray}
for $k=1,2,3$. This and the reinterpreted equations (\ref{eq:map1}) define a linear map of $4 \times 4$ matrices to $4 \times 4$ matrices. If the map of $2 \times 2$ matrices defined by equations  (\ref{eq:map1}) is completely positive, this map of $4 \times 4$ matrices should take every positive matrix to a positive matrix. We will see that it can fail to do that even when the $4 \times 4$ matrix being mapped is a density matrix for a possible initial state of the two qubits.

If $\Pi$ is a density matrix for the two qubits then
\begin{equation}
  \label{eq:density4}
  \Pi = \frac{1}{4} \left( 1 + \sum_{j=1}^{3} \apsig{j}\psig j + \sum_{k=1}^{3} \apxi{k}\pxi k + \sum_{j,k=1}^{3} \apsigxi{j}{k} \psig j \pxi k \right)
\end{equation}
is mapped to
\begin{equation}
  \label{eq:density4a}
  \Pi' = \frac{1}{4} \left( 1' + \sum_{j=1}^{3} \apsig{j}\psig j' + \sum_{k=1}^{3} \apxi{k} \pxi k' + \sum_{j,k=1}^{3} \apsigxi{j}{k} (\psig j \pxi k)' \right).
\end{equation}
To test whether $\Pi'$ is positive, let
\begin{equation}
  \label{eq:pos5}
  W = \frac{1}{4} \left( 1 + \frac{1}{\sqrt{2}} \psig2 + \frac{1}{\sqrt{2}} \psig 3 \pxi 3 \right),
\end{equation} 
check that $W^2 = \frac{1}{2} W$ to see that $W$ is positive and is a density matrix, and calculate
\begin{equation}
  \label{eq:pos6}
  {\mbox{Tr}}[\Pi' W] = \frac{1}{4} \left( 1 + \frac{a_2}{\sqrt{2}} + \frac{\apsigxi{3}{3}}{\sqrt{2}} \right)  = \frac{1}{4} \left( 1 + \frac{\apsigxi{1}{1}}{\sqrt{2}} + \frac{\apsigxi{3}{3}}{\sqrt{2}} \right).
\end{equation}
This holds if $\Pi$ is the density matrix for an initial state of the two qubits that gives the mean values $-\apsigxi{2}{1}$ and $\apsigxi{1}{1}$ used for $a_1$ and $a_2$. We see that ${\mbox{Tr}}[\Pi' W]$ can be negative. Both $\apsigxi{1}{1}$ and $\apsigxi{3}{3}$ are $-1$ for the state where the sum of the spins of the two qubits is zero. That state gives zero for $\avecsig$, but nearby states will give an acceptable set of compatible $\avecsig$ with  ${\mbox{Tr}}[\Pi' W]$ negative. 

The map is made to be used for the set of states, the set of density matrices, described by compatible $\avecsig$. We call that its {\em compatibility domain}. It includes all the initial states the $\Sigma$ qubit can have with the given $\apsigxi{2}{1}$ and $\apsigxi{1}{1}$. Outside the compatibility domain, some density matrices are mapped to positive matrices, but others, including, for example, $P$ from equation (\ref{eq:op1}), are not. Even inside its compatibility domain, the map is not completely positive.

We can see that the compatibility domain is enough to give the linearity of the map physical meaning. Applied to density matrices, the linearity of the map says that if density matrices $\rho$ and $\sigma$ are mapped to $\rho'$ and $\sigma'$ then each density matrix 
\begin{equation}
  \label{eq:density7}
  \tau = q \rho + ( 1-q)\sigma
\end{equation}
with $0 < q < 1$ is mapped to 
\begin{equation}
  \label{eq:density8}
  \tau' = q \rho' + (1- q) \sigma' .
\end{equation}
Suppose $\rho$ and $\sigma$ are density matrices for the $\Sigma$ qubit that give mean values $\avecsig _{\rho}$ and $\avecsig _{\sigma}$. If both  $\avecsig _{\rho}$ and $\avecsig _{\sigma}$ are compatible with the same $\apsigxi{2}{1}$ and $\apsigxi{1}{1}$ in describing an initial state of the two qubits, then so is
\begin{equation}
  \label{eq:density9}
  \avecsig _{\tau} = q \avecsig _{\rho} + (1- q) \avecsig _{\sigma} .
\end{equation}
The compatibility domain is convex. Explicitly, if $\Pi_{\rho}$ and $\Pi_{\sigma}$ are density matrices for the two qubits written in the form of equation (\ref{eq:density4}) with $\avecsig _{\rho}$ and $\avecsig_{\sigma}$ for $\avecsig$ and the same $\apsigxi{2}{1}$ and $\apsigxi{1}{1}$, then
\begin{equation}
  \label{eq:density10}
  \Pi_{\tau} = q \Pi_{\rho} + (1-q) \Pi_{\sigma}
\end{equation}
is a density matrix for the two qubits written in the same form with $\avecsig _{\tau}$ for $\avecsig$ and the  same $\apsigxi{2}{1}$ and $\apsigxi{1}{1}$. If $\rho$ and $\sigma$ are in the compatibility domain, then so are all the $\tau$ defined by equation (\ref{eq:density7}). For these, the linearity described by equations (\ref{eq:density7}) and (\ref{eq:density8}) has a meaningful physical interpretation. The compatibility domain will be described more completely in Section \ref{domains}.

A different map is an option if the initial state of the two qubits is a product state or if at least
\begin{equation}
  \label{eq:prod1}
  \apsigxi{2}{1} = \apsig{2} \apxi{1} \, , \qquad \apsigxi{1}{1} = \apsig{1} \apxi{1} .
\end{equation}
Then the density matrix $\rho'$ described by equation (\ref{eq:density2}) is 
\begin{equation}
  \label{eq:density11}
  \rho' = \frac{1}{2} ( 1 - \apsig{2} \apxi{1} \psig1 + \apsig{1} \apxi{1} \psig2 + \apsig{3} \psig3).
\end{equation}
This is obtained from equation (\ref{eq:density3}) with the linear map of $2 \times 2$ matrices defined either by equations (\ref{eq:map1}) or by 
\begin{equation}
  \label{eq:map2}
  1'=1 \, , \qquad \psig1' = \apxi{1} \psig2 \, , \qquad \psig2' = - \apxi{1} \psig1 \, , \qquad \psig3' = \psig3 .
\end{equation}
With the latter, every positive matrix maps to a positive matrix. In fact the map is completely positive.

This completely positive map is defined by equations (\ref{eq:map2}) for a given fixed value of $\apxi{1}$. That puts no restrictions on $\avecsig$, no limits on the initial state of the $\Sigma$ qubit. Every $\avecsig$ is compatible with any $\apxi{1}$ in describing an initial state of the two qubits for which equations (\ref{eq:prod1}) hold; every state of the $\Sigma$ qubit can be combined with any state of the $\Xi$ qubit in a product state for the two qubits. However we will see that the $\avecsig$ compatible with given nonzero $\apsigxi{2}{1}$ and $\apsigxi{1}{1}$ in product states for the two qubits fill only a two-dimensional set embedded in the three-dimensional compatibility domain.

The completely positive map defined by equations (\ref{eq:map2}) is an option only when equations (\ref{eq:prod1}) hold. Then both maps, from equations (\ref{eq:map1}) and (\ref{eq:map2}), reproduce the evolution of the $\Sigma$ qubit. There is a map defined by equations (\ref{eq:map1}) for almost every initial state of the two qubits, with $\apsigxi{2}{1}$ and $\apsigxi{1}{1}$ changing continuously from state to state. Switching to the completely positive map when equations (\ref{eq:prod1}) hold would be a discontinuous change. 

\subsection{Time dependence}\label{sec2b}

From the mean values (\ref{eq:evol1}) for any $t$, the same steps as before with equations (\ref{eq:density1}), (\ref{eq:density2}) and (\ref{eq:density3}) yield
\begin{eqnarray}
  \label{eq:anyt1}
  1' = 1+ (a_1 \psig1 + a_2 \psig2) \sin \omega t \hspace{2 cm}\nonumber \\
\psig1' = \psig1 \cos \omega t \, , \qquad \psig2' = \psig2 \cos \omega t \, , \qquad \psig3 ' = \psig3 .
\end{eqnarray}
This defines a linear map $Q \rightarrow Q'$ of all $2 \times 2$ matrices to $2 \times 2$ matrices described by
\begin{equation}
  \label{eq:bmatrix1}
  Q_{rs}'= \sum_{jk} {\cal{B}}_{rj; sk} Q_{jk}  
\end{equation}
with
\begin{equation}
  \label{eq:bmatrix2}
  {\cal{B}} = \left( \begin{array}{cccc}
 1 & 0 & \frac{1}{2} a^* \sin \omega t & \cos \omega t \\
 0 & 0 & 0 &  \frac{1}{2} a^* \sin \omega t \\
  \frac{1}{2} a \sin \omega t & 0 & 0 & 0 \\
\cos \omega t &  \frac{1}{2} a \sin \omega t & 0 & 1 \end{array} \right)
\end{equation}
where $a=a_1 + ia_2$ and the rows and columns of ${\cal{B}}$ are in the order $11$, $12$, $21$, $22$.

A vector
\begin{equation}
  \label{eq:beig1}
  \psi_{1 \; {\mbox{or}} \; 3} = \left( \begin{array}{c} \lambda \\ \frac{1}{2} a^* \sin \omega t  \\   \frac{1}{2} a \sin \omega t \\ \lambda \end{array} \right) 
\end{equation}
is an eigenvector of ${\cal{B}}$ with eigenvalue $\lambda$ if
\begin{equation}
  \label{eq:eigeq1}
  \lambda + \frac{1}{4} |a|^2 \sin^2 \omega t + \lambda \cos \omega t = \lambda^2 .
\end{equation}
This yields two eigenvalues
\begin{eqnarray}
  \label{eq:beig2}
  \lambda_1 & = & \frac{1}{2} \left( 1 + \cos \omega t + \sqrt{(1 + \cos \omega t)^2 + |a|^2 \sin^2 \omega t} \right) \nonumber \\
  \lambda_3 & = & \frac{1}{2} \left( 1 + \cos \omega t - \sqrt{(1 + \cos \omega t)^2 + |a|^2 \sin^2 \omega t} \right)
\end{eqnarray}
and eigenvectors 
\begin{eqnarray}
  \label{eq:beig3}
   \psi_{1 \; {\mbox{or}} \; 3} & = & \psi_1 \quad {\mbox{for}} \quad \lambda = \lambda_1 \nonumber \\
 & = & \psi_3 \quad {\mbox{for}} \quad \lambda = \lambda_3 .
\end{eqnarray}
Note that $\psi_1$ and $\psi_3$ are orthogonal because
\begin{equation}
  \label{eq:orth1}
  \lambda_1 \lambda_3 =- \frac{1}{4} |a|^2 \sin^2 \omega t.
\end{equation}
The squares of the lengths of the eigenvectors are
\begin{eqnarray}
  \label{eq:beig4}
  ||\psi_n||^2& =& 2\left(\lambda_n^2 + \frac{1}{4} |a|^2 \sin^2 \omega t \right) \nonumber \\
& = & 2 \lambda_n ( 1 + \cos \omega t) + |a|^2 \sin^2 \omega t
\end{eqnarray}
for $n=1,3.$
A vector
\begin{equation}
  \label{eq:beig1a}
  \psi_{2 \; {\mbox{or}} \; 4} = \left( \begin{array}{c} \lambda \\ -\frac{1}{2} a^* \sin \omega t  \\   \frac{1}{2} a \sin \omega t \\ -\lambda \end{array} \right) 
\end{equation}
is an eigenvector of ${\cal{B}}$ with eigenvalue $\lambda$ if
\begin{equation}
  \label{eq:eigeq1a}
  \lambda + \frac{1}{4} |a|^2 \sin^2 \omega t - \lambda \cos \omega t = \lambda^2.   
\end{equation}
This yields two eigenvalues
\begin{eqnarray}
  \label{eq:beig2a}
  \lambda_2 & = & \frac{1}{2} \left( 1 - \cos \omega t + \sqrt{(1 - \cos \omega t)^2 + |a|^2 \sin^2 \omega t} \right) \nonumber \\
  \lambda_4 & = & \frac{1}{2} \left( 1 - \cos \omega t - \sqrt{(1 - \cos \omega t)^2 + |a|^2 \sin^2 \omega t} \right)
\end{eqnarray}
and eigenvectors 
\begin{eqnarray}
  \label{eq:beig3a}
   \psi_{2 \; {\mbox{or}} \; 4} & = & \psi_2 \quad {\mbox{for}} \quad \lambda = \lambda_2 \nonumber \\
 & = & \psi_4 \quad {\mbox{for}} \quad \lambda = \lambda_4 .
\end{eqnarray}
Note that $\psi_2$ and $\psi_4$ are orthogonal because
\begin{equation}
  \label{eq:orth1a}
  \lambda_2 \lambda_4 = -\frac{1}{4} |a|^2 \sin^2 \omega t.
\end{equation}
The squares of the lengths of the eigenvectors are
\begin{eqnarray}
  \label{eq:beig4a}
  ||\psi_n||^2& =& 2\left(\lambda_n^2 + \frac{1}{4} |a|^2 \sin^2 \omega t \right) \nonumber \\
& = & 2 \lambda_n ( 1 - \cos \omega t) + |a|^2 \sin^2 \omega t
\end{eqnarray}
for $n=2,4.$

\begin{figure}[htb]
   \begin{center}
     \includegraphics{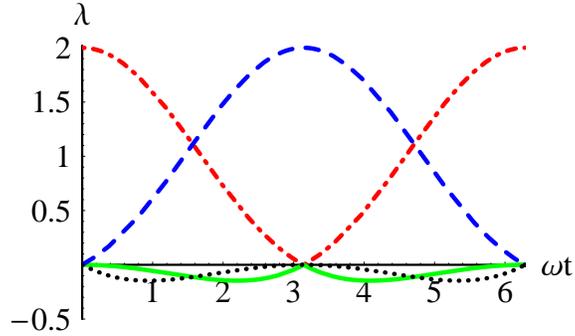}
   \end{center}
   \caption{(color online) The eigenvalues of ${\cal{B}}$ as a function of $\omega t$ when $|a|^2$ is $\frac{1}{2}$. The dot-dash (red) line is $\lambda_1$, the solid (green) line is $\lambda_2$, the dashed (blue) line is $\lambda_3$, and dotted (black) line  is $\lambda_4$. }
\label{fig1}
\end{figure}

We see that, in all but a few exceptional cases, ${\cal{B}}$ has two positive eigenvalues $\lambda_1$ and $\lambda_2$ and two negative eigenvalues $\lambda_3$ and $\lambda_4$. That means the map is not completely positive; for a completely positive map, ${\cal{B}}$ is a positive matrix and its eigenvalues are all non-negative. A plot of the eigenvalues of ${\cal{B}}$ as a function of $\omega t$ when $|a|^2$ is $1/2$ is shown in Figure \ref{fig1}.
The two negative eigenvalues $\lambda_3$ and $\lambda_4$ go to zero when $\omega t$ is $n \pi$; the map is the identity map for even $n$ and rotation by $\pi$ around the $z$ axis for odd $n$.  

The spectral decomposition 
\begin{equation}
  \label{eq:bspec1}
  {\cal{B}} = \sum_{n=1}^4 \lambda_n | n \rangle \langle n|
\end{equation}
with
\begin{equation}
  \label{eq:bspec2}
  |n \rangle = \frac{1}{||\psi_n||} | \psi_n \rangle
\end{equation}
yields
\begin{equation}
  \label{eq:bsepc3}
  B_{rj;sk} = \sum_{n=1}^4 \lambda_n \langle r \, j | n \rangle \langle s \, k | n \rangle^* = \sum_{n=1}^4{\mbox{sign}}(\lambda_n) C(n)_{rj} C(n)_{ks}^{\dagger} 
\end{equation}
with
\begin{equation}
  \label{eq:cmatrix1}
  C(n)_{rj} = \sqrt{|\lambda_n|} \langle r \, j | n \rangle = \frac{\sqrt{|\lambda_n | }}{||\psi_n||} \langle r \, j | \psi_n \rangle
\end{equation}
so equation (\ref{eq:bmatrix1}) is
\begin{equation}
  \label{eq:cmatrix2}
  Q'_{rs} = \sum_{n=1}^4 {\mbox{sign}} (\lambda_n) \sum_{jk} C(n)_{rj} Q_{jk} C(n)_{ks}^{\dagger}
\end{equation}
or
\begin{equation}
  \label{eq:cmatri3}
  Q' = \sum_{n=1}^4 {\mbox{sign}}(\lambda_n) C(n) Q C(n)^{\dagger}. 
\end{equation}
Since ${\mbox{Tr}}Q' = {\mbox{Tr}}Q$ for all $Q$ for our map, 
\begin{equation}
  \label{eq:cmatrix4}
  \sum_{n=1}^4 {\mbox{sign}}(\lambda_n) C(n)^{\dagger} C(n) = 1.
\end{equation}
Except for the minus signs, these equations are the same as for completely positive maps. Explicitly we have
\begin{equation}
  \label{eq:cmatrix5}
  C(n) = \sqrt{\frac{|\lambda_n|}{2 \lambda_n (1 + \cos \omega t) + |a|^2 \sin^2 \omega t}} \left[ \lambda_n + \frac{1}{2} ( a_1 \psig1 + a_2 \psig2) \sin \omega t \right] 
\end{equation}
for $n = 1, 3$, and 
\begin{equation}
  \label{eq:cmatrix6}
  C(n) = \sqrt{\frac{|\lambda_n|}{2 \lambda_n (1 - \cos \omega t) + |a|^2 \sin^2 \omega t}} \left[ \lambda_n \psig3+ \frac{i}{2} ( a_2 \psig1 - a_1 \psig2) \sin \omega t \right] 
\end{equation}
for $n=2,4$. For small $\omega t$ and nonzero $|a|$
\begin{eqnarray}
  \label{eq:lambdasmall}
  \lambda_1 & = & 2 - \frac{1}{2} ( \omega t)^2 + \frac{1}{8} |a|^2 (\omega t)^2 \nonumber \\
\lambda_2 & =& \frac{1}{2} | a| \omega t + \frac{1}{4} (\omega t)^2 + \frac{1}{16 |a|} (\omega t)^3 \nonumber \\
\lambda_3& = &- \frac{1}{8} |a|^2 (\omega t)^2  \nonumber \\
\lambda_4& =& - \frac{1}{2} | a| \omega t + \frac{1}{4} (\omega t)^2 - \frac{1}{16 |a|} (\omega t)^3 
\end{eqnarray}
and
\begin{eqnarray}
  \label{eq:cmatrixsmall}
  C(1)& = &1 - \frac{(\omega t)^2}{8} + \frac{\omega t}{4} ( a_1 \psig1 + a_2 \psig2) \nonumber \\
C(2)& =& \sqrt{\frac{|a|}{8}} \left[ (\omega t)^{\frac{1}{2}} + \frac{1}{2 |a|} (\omega t)^{\frac{3}{2} }\right] \psig3 + \sqrt{\frac{1}{8 |a|}} (\omega t)^{\frac{1}{2}} (i a_2 \psig1 - i a_1 \psig2) \nonumber \\
  C(3)& = &- \frac{|a|^2}{16}(\omega t)^2 + \frac{\omega t}{4} ( a_1 \psig1 + a_2 \psig2) \nonumber \\
C(4)& =& \sqrt{\frac{|a|}{8}} \left[- (\omega t)^{\frac{1}{2}} + \frac{1}{2 |a|} (\omega t)^{\frac{3}{2}} \right] \psig3 + \sqrt{\frac{1}{8 |a|}} (\omega t)^{\frac{1}{2}} (i a_2 \psig1 - i a_1 \psig2).
\end{eqnarray}

\subsection{Compatibility and positivity domains}\label{domains}

Now we describe the compatibility and positivity domains completely and precisely. To write equations for the compatibility domain, we make a convenient choice of components for $\avecsig$. Suppose $a_1$ and $a_2$ are given. Then $\apsigxi{1}{1}$ and $\apsigxi{2}{1}$ are fixed. Let
\begin{eqnarray}
  \label{eq:domain1}
  \psig+ &=& \frac{\apsigxi{1}{1} \psig1 + \apsigxi{2}{1}\psig2}{\sqrt{\apsigxi{1}{1}^2 + \apsigxi{2}{1}^2 }} \nonumber \\
 \psig- &=& \frac{\apsigxi{2}{1} \psig1 - \apsigxi{1}{1}\psig2}{\sqrt{\apsigxi{1}{1}^2 + \apsigxi{2}{1}^2 }}.
\end{eqnarray}
Then $\psig+$ and $\psig-$ anticommute, their squares are both $1$, and $\apsigxi{-}{1}$ is zero,
\begin{equation}
  \label{eq:domain2}
  \apsigxi{+}{1} = \sqrt{\apsigxi{1}{1}^2 + \apsigxi{2}{1}^2 }= \sqrt{(a_1)^2 + (a_2)^2},
\end{equation}
and
\begin{equation}
  \label{eq:domain3}
  \apsigxi{1}{1} \psig1 \pxi1 + \apsigxi{2}{1} \psig2 \pxi1 = \apsigxi{+}{1} \psig+ \pxi1. 
\end{equation}
The compatibility domain is the set of $\avecsig$, or $\apsig{+}$, $\apsig{-}$, $\apsig{3}$, that are compatible with the given $\apsigxi{+}{1}$ and zero $\apsigxi{-}{1}$ in describing a possible initial state for the two qubits. 

\begin{figure}[htb]
   \begin{center}
   \includegraphics{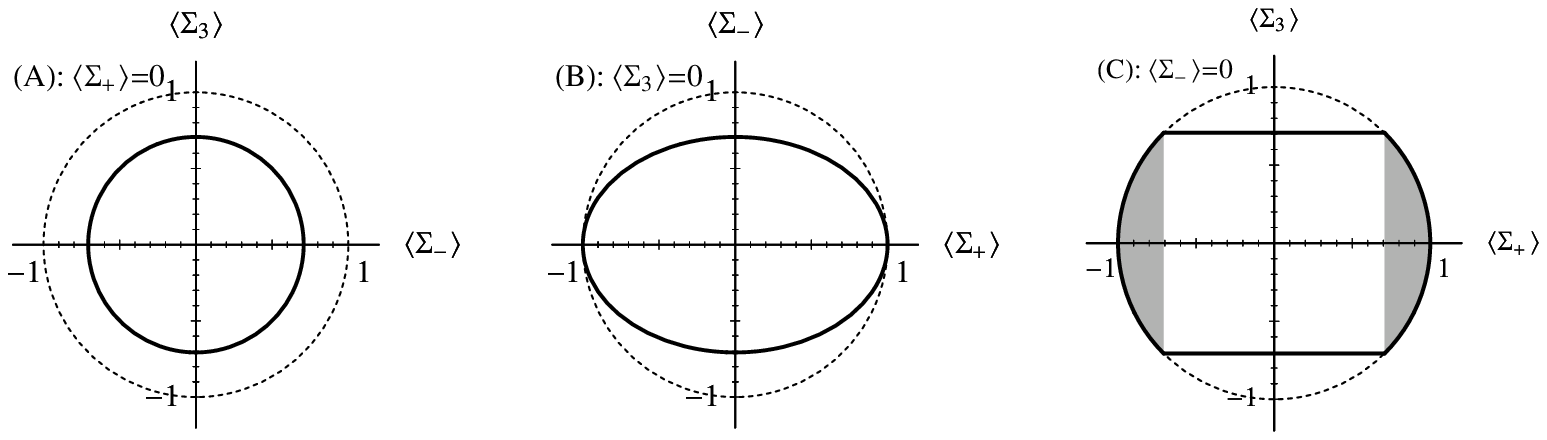}
   \end{center}
   \caption{Sections of the compatibility domain when $\apsigxi{+}{1}=\frac{1}{\sqrt{2}}$. The area enclosed by the thick solid line is the compatibilty domain. The dotted line shows the unit circle. The shaded area in (C) shows the $\avecsig$ for product states compatible with the given $\apsigxi{+}{1}$ and zero $\apsigxi{-}{1}$.}
\label{fig2}
\end{figure}

Basic outlines of the compatibility domain are easy to see. When $\apsig{+}$ is zero, the compatibility domain includes the $\apsig{-}$, $\apsig{3}$ such that
\begin{equation}
  \label{eq:domain4}
  \apsig{-}^2 + \apsig{3}^2 + \apsigxi{+}{1}^2 \leq 1
\end{equation}
because for these
\begin{equation}
  \label{eq:domain5}
  \Pi = \frac{1}{4} ( 1 + \apsig{-} \psig- + \apsig{3} \psig3 + \apsigxi{+}{1} \psig+ \pxi1) 
\end{equation}
is a density matrix for the two qubits. Larger $\apsig{-}$ and $\apsig{3}$ are not included. If 
\begin{equation}
\label{eq:condition1}
(x_-)^2+(x_3)^2+\apsigxi{+}{1}^2=1
\end{equation}
and $r > 1$, then
\begin{eqnarray}
  \label{eq:domain6}
  \Pi & =& \frac{1}{4}( 1 + r x_{-}\psig- + r x_3 \psig3 + \apsigxi{+}{1} \psig+ \pxi1 \nonumber \\
&& \hspace{1 cm} + \sum_{j=1}^{3} y_j \pxi{j} + z_{31} \psig3 \pxi1 + \sum_{j=1}^{3} \sum_{k=2}^{3} z_{jk} \psig{j} \pxi{k})
\end{eqnarray}
is not a density matrix for any $y_j$ and $z_{jk}$ because 
\begin{equation}
  \label{eq:domain7}
  W = \frac{1}{4} ( 1 - x_{-} \psig- - x_3 \psig3 - \apsigxi{+}{1}\psig+ \pxi1)
\end{equation}
is a density matrix and
\begin{equation}
  \label{eq:domain8}
  {\mbox Tr}[\Pi W] = \frac{1}{4} ( 1 - r (x_{-})^2 - r (x_3)^2 - \apsigxi{+}{1}^2 ) < 0.
\end{equation}
When $\apsig{+}$ is zero, the compatibility domain is just the circular area described by (\ref{eq:domain4}); it cannot be extended in any direction described by any ratio of $\apsig{-}$ and $\apsig{3}$. This projection of the compatibility domain on the $\apsig{-}$, $\apsig{3}$ plane is shown in Figure \ref{fig2}-A for the case where $\apsigxi{+}{1}$ is $1/\sqrt{2}$.

When $\apsig{3}$ is zero, the compatibility domain is the elliptical area of $\apsig{-}$, $\apsig{+}$ such that
\begin{equation}
	\label{eq:domain20}
	\frac{\apsig{-}^2}{1-\apsigxi{+}{1}^2} + \apsig{+}^2 \leq 1.
\end{equation}
To see this, we find when the eigenvalues of 
\begin{equation}
	\label{eq:domain21}
	\Pi = \frac{1}{4} ( 1 + \apsig{-} \psig- + \apsig{3}\psig3 + \apsig{+} \psig+ + \apsigxi{+}{1} \psig+ \pxi1 + \apxi{1} \pxi1 + \apsigxi{3}{1} \psig3 \pxi1)
\end{equation}
are all nonnegative so that $\Pi$ is a density matrix for the two qubits. Let
\begin{equation}
\Pi = \frac{1}{4} ( 1 + \apxi{1}\pxi1 + M ). 
\end{equation}
Then
\begin{equation}
	M^2 = \apsig{-}^2 + \apsig{3}^2 + \apsig{+}^2 + \apsigxi{+}{1}^2 + \apsigxi{3}{1}^2 + 2 \apsig{3}\apsigxi{3}{1}\pxi1 + 2 \apsig{+}\apsigxi{+}{1} \pxi1.
\end{equation}
The eigenvalues of $M$ are the square roots of the eigenvalues of $M^2$. When $\pxi1$ has eigenvalue $+1$, the eigenvalues of $\Pi$ are
\begin{equation}
	\frac{1}{4} ( 1 + \apxi{1} \pm \sqrt{m^2(+)})
\end{equation}
where $m^2(+)$ is $M^2$ with $\pxi1$ replaced by its eigenvalue $+1$. When $\pxi1$ has eigenvalue $-1$, the eigenvalues of $\Pi$ are
\begin{equation}
	\frac{1}{4} ( 1 - \apxi{1} \pm \sqrt{m^2(-)})
\end{equation}
where $m^2(-)$ is $M^2$ with $\pxi1$ replaced by its eigenvalue $-1$. The eigenvalues of $\Pi$ are all nonnegative if
\begin{equation}
	\label{eq:domain22}
	m^2(+) \leq ( 1 + \apxi{1})^2 
\end{equation}
\begin{equation}
	\label{eq:domain23}
	m^2(-) \leq ( 1 - \apxi{1})^2
\end{equation}
and $\apxi{1}^2 \leq 1$. When $\apsig{3}$ is zero, the areas of $\apsig{-}$, $\apsig{+}$ allowed by the inequalities (\ref{eq:domain22}) and (\ref{eq:domain23}) are largest when $\apsigxi{3}{1}$ is zero. Then as $\apxi{1}$ varies from $-1$ to $1$ the inequalities (\ref{eq:domain22}) and (\ref{eq:domain23}) describe the area of an ellipse with foci at $\pm \apsigxi{+}{1}$ on the $\apsig{+}$ axis; they say that the distance from a point with coordinates $\apsig{-}$, $\apsig{+}$ to the focus at $-\apsigxi{+}{1}$ is bounded by $1 + \apxi{1}$ and the distance to the focus at $\apsigxi{+}{1}$ is bounded by $1-\apxi{1}$, so the sum of the distances is bounded by $2$. That gives the elliptical area described by (\ref{eq:domain20}). We conclude that it is the compatibility domain when $\apsig{3}$ is zero. This conclusion is not changed if $\Pi$ is given additional terms involving $\pxi2$, $\pxi3$, $\psig{j} \pxi2$, $\psig{j} \pxi3$. Each eigenvalue that we considered is a diagonal matrix element $(\psi, \Pi \psi)$ with $\psi$ and eigenvector of $\pxi1$ as well as an eigenvector of the $\Pi$ we considered, so $(\psi, \pxi2 \psi)$, $(\psi, \pxi3 \psi)$, $(\psi, \psig{j}\pxi2 \psi)$, $(\psi, \psig{j} \pxi3 \psi)$ are zero. Additional terms will change the eigenvalues and eigenvectors of $\Pi$ but will not change the diagonal matrix elements we considered. They have to be nonnegative if $\Pi$ is a density matrix. That is all we need to show that the inequality (\ref{eq:domain20}) describes the compatibility domain when $\apsig{3}$ is zero. The projection of the compatibility domain on the $\apsig{+}$, $\apsig{-}$ plane is shown in Figure \ref{fig2}-B for the case where $\apsigxi{+}{1}$ is $1/\sqrt{2}$.

When $a_1$ and $a_2$ are not both zero, all the product states for the two qubits that are compatible with the given $\apsigxi{+}{1}$ and zero $\apsigxi{-}{1}$ are for $\avecsig$ in the projection of the compatibility domain in the $\apsig{3}$, $\apsig{+}$ plane. If
\begin{eqnarray}
	\apsig{-}\apxi{1} &=& \apsigxi{-}{1}=0 \nonumber \\
	\apsig{+}\apxi{1} &=& \apsigxi{+}{1} \neq 0
\end{eqnarray}
then $\apsig{-}=0$ and
\begin{equation}
	\label{eq:domain24}
	\apsig{+}^2 \geq \apsigxi{+}{1}^2.
\end{equation}
There is a compatible product state for each such $\apsig{+}$ and each $\apsig{3}$ such that
\begin{equation}
	\label{eq:domain25}
	\apsig{3}^2 \leq 1-\apsig{+}^2,
\end{equation}
with $\apsig{-}=0$. The $\avecsig$ for compatible product states fill the two areas in the $\apsig{+}$, $\apsig{3}$ plane bounded by sections of the unit circle from (\ref{eq:domain25}) and  straight lines from (\ref{eq:domain24}). These areas are shown in Figure \ref{fig2}-C for the case where $\apsigxi{+}{1}$ is $1/\sqrt{2}$.

Since $\avecsig$ cannot be outside the unit circle for any state, these sections of the unit circle are on the boundary of the compatibility domain. We can conclude that the boundary of the projection of the compatibility domain in the $\apsig{+}$, $\apsig{3}$ plane is completed by straight lines with constant values of $\apsig{3}$ between the sections of the unit circle, because we proved the compatibility domain is convex and from (\ref{eq:domain4}), (\ref{eq:domain25}) and (\ref{eq:domain24}) we see that $\apsig{3}^2$ cannot be larger when $\apsig{+}$ is zero then it is at the termini of the sections of the unit circle. The complete boundary in shown in Figure \ref{fig2}-C for the case where $\apsigxi{+}{1}$ is $1/\sqrt{2}$.

We will show that the compatibility domain is the set of $\avecsig$ where
\begin{equation}
	\label{eq:domain26}
	\sqrt{(\apsig{-}^2 + \apsig{+}^2 +\apsigxi{+}{1}^2)^2-4\apsig{+}^2 \apsigxi{+}{1}^2} \leq 2 - 2 \apsig{3}^2 - \apsig{-}^2 -\apsig{+}^2- \apsigxi{+}{1}^2.  
\end{equation}
First let us see what this says. Squaring both sides of (\ref{eq:domain26}) gives
\begin{equation}
	\label{eq:domain27}
	\apsig{-}^2 + \apsig{+}^2 + \apsig{3}^2 + \apsigxi{+}{1}^2- \frac{\apsig{+}^2 \apsigxi{+}{1}^2}{1 - \apsig{3}^2} \leq 1.
\end{equation}
When $\apsig{+}$ is zero, (\ref{eq:domain27}) is the inequality (\ref{eq:domain4}) that describes the circular projection of the compatibility domain in the $\apsig{-}$, $\apsig{3}$ plane. When $\apsig{3}$ is zero, (\ref{eq:domain27}) is the inequality (\ref{eq:domain20}) that describes the elliptical projection of the compatibility domain in the $\apsig{-}$, $\apsig{+}$ plane. If $\apsig{3}^2$ is between zero and $1-\apsigxi{+}{1}^2$, then (\ref{eq:domain27}) is
\begin{equation}
	\label{eq:domain28}
	\frac{\apsig{-}^2}{1 - \apsigxi{+}{1}^2 - \apsig{3}^2}+ \frac{\apsig{+}^2}{1-\apsig{3}^2} \leq 1.
\end{equation}
A contour of the compatibility domain at constant $\apsig{3}$ is an ellipse. As $\apsig{3}^2$ approaches $1-\apsigxi{+}{1}^2$ the semi-minor axis shrinks to zero and the semi-major axis goes to $\apsigxi{+}{1}$, so the ellipse reduces to a line from $-\apsigxi{+}{1}$ to $\apsigxi{+}{1}$ along the $\apsig{+}$ axis. When $\apsig{-}$ is zero, (\ref{eq:domain26}) is 
\begin{equation}
	\label{eq:domain29}
	\apsig{3}^2 \leq 1 - \frac{\apsig{+}^2 + \apsigxi{+}{1}^2}{2} - \frac{|\apsig{+}^2 - \apsigxi{+}{1}^2|}{2}
\end{equation}
which is (\ref{eq:domain25}) when $\apsig{+}^2 \geq \apsigxi{+}{1}^2$ and is 
\begin{equation}
	\label{eq:domain30}
	\apsig{3}^2 \leq 1 - \apsigxi{+}{1}^2 
\end{equation}
when $\apsig{+}^2 \leq \apsigxi{+}{1}^2$. That describes the area bounded by sections of the unit circle and straight lines that is the projection of the compatibility domain in the $\apsig{+}$, $\apsig{3}$ plane. 

\begin{figure}[htb]
   \begin{center}
   \includegraphics{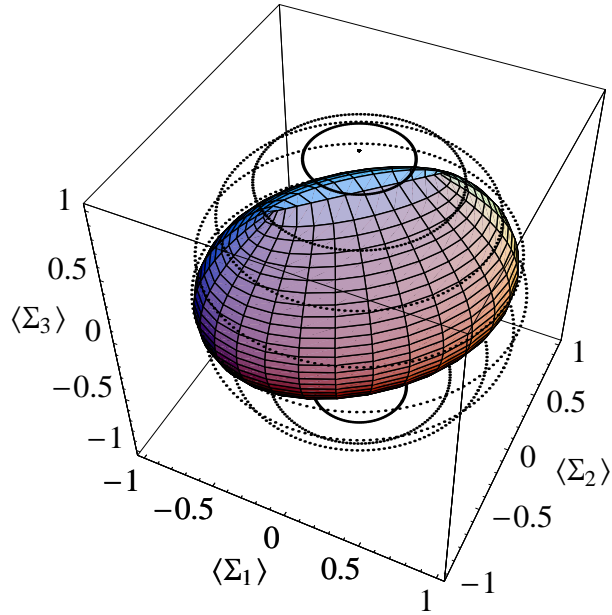}
   \end{center}
   \caption{(color online) The compatibility domain generated using {\em Mathematica} for the case where $\apsigxi{2}{1}$ and $\apsigxi{1}{1}$ are both $\frac{1}{2}$. The dotted sphere is the unit sphere (the Bloch sphere) that represents all possible states of the qubit.}
\label{fig2a}
\end{figure}

When $\apsigxi{+}{1}$ is zero, (\ref{eq:domain26}) just says that $\avecsig$ is on or inside the unit sphere; then there is no restriction on $\avecsig$ from compatibility. A three-dimensional view of the compatibility domain is shown in Figure \ref{fig2a} for the case where $\apsigxi{+}{1}$ is $1/\sqrt{2}$.
 
The inequality (\ref{eq:domain26}) puts a bound on $\apsig{3}^2$ for each $\apsig{-}$ and $\apsig{+}$. In particular, it says $\apsig{3}^2$ can never be larger than the values it has when $\apsig{-}$ is zero; the bound (\ref{eq:domain30}) holds for the entire compatibility domain. For $\apsig{3}^2$ within this bound, the left side of (\ref{eq:domain27}) is an increasing function of $\apsig{+}^2$. The inequality (\ref{eq:domain27}) puts a bound on $\apsig{-}^2$ for each $\apsig{+}$ and $\apsig{3}$ and a bound on $\apsig{+}^2$ for each $\apsig{-}$ and $\apsig{3}$. 

To show that the set of $\avecsig$ described by (\ref{eq:domain26}) is in the compatibility domain, we show that for each $\avecsig$ that satisfies (\ref{eq:domain26}) there is a $\Pi$ described by (\ref{eq:domain21}) that is a density matrix for the two qubits. We let
\begin{equation}
	\label{eq:domain31}
	\apxi{1} = \frac{\apsig{+}\apsigxi{+}{1}}{1 - \apsig{3}^2}
\end{equation}
and
\begin{equation}
	\label{eq:domain32}
	\apsigxi{3}{1}=\apsig{3}\apxi{1}.
\end{equation}
Then the inequalities (\ref{eq:domain22}) and (\ref{eq:domain23}) are both (\ref{eq:domain27}). From (\ref{eq:domain30}), which (\ref{eq:domain26}) implies, 
\begin{equation}
	\label{eq:domain33}
	|\apxi{1}| \leq \frac{|\apsig{+}|}{\apsigxi{+}{1}} \leq 1
\end{equation}
for $\apsig{+}^2 \leq \apsigxi{+}{1}^2$, and from (\ref{eq:domain25}), which holds for any $\avecsig$,
\begin{equation}
	|\apxi{1}| \leq \frac{\apsigxi{+}{1}}{|\apsig{+}|} \leq 1
\end{equation}
 for $\apsig{+}^2 \geq \apsigxi{+}{1}^2$. This implies that the eigenvalues of $\Pi$ are all nonnegative, which means $\Pi$ is a density matrix for the two qubits. 
  
 The inequality (\ref{eq:domain27}) by itself does not imply that $\avecsig$ is in the compatibility domain. The equality limit of (\ref{eq:domain27}) is a quadratic equation for $\apsig{3}^2$. The equality limit of (\ref{eq:domain26}) is one solution. In the other solution, the sign of the square root in (\ref{eq:domain26}) is changed. That changes the sign of the term with the absolute value in (\ref{eq:domain29}), which extends the boundary to include the entire area of the unit circle in the $\apsig{+}$, $\apsig{3}$ plane. The bounds (\ref{eq:domain30}) on $\apsig{3}^2$ and (\ref{eq:domain33}) on $|\apxi{1}|$ do not hold for the other solution. They are not implied by (\ref{eq:domain26}).

We have shown that the set of $\avecsig$ described by the inequality (\ref{eq:domain26}) is in the compatibility domain. The compatibility domain is the same for all $t$. In a larger domain, which we call the {\em positivity domain}, every positive matrix is mapped to a positive matrix. The positivity domain depends on the time $t$. We will show that the set of $\avecsig$ described by the inequality (\ref{eq:domain26}) is also the intersection of all the positivity domains for different $t$. That implies it is the compatibility domain; the compatibility domain cannot be larger, because it must be in every positivity domain for every $t$.

\begin{figure}[htb]
   \begin{center}
   \includegraphics{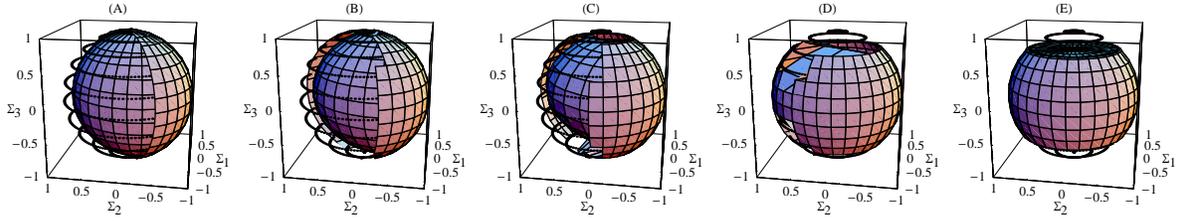}
   \end{center}
   \caption{(color online) The positivity domains for (left to right)  $\omega t = \frac{\pi}{10}$, $\frac{2\pi}{10}$, $\frac{3 \pi}{10}$, $\frac{4 \pi}{10}$ and $\frac{ \pi}{2}$ when $a_1$ is $-\frac{1}{2}$ and $a_2$ is $\frac{1}{2}$. The surface of the unit sphere is shown with dotted lines where it is not the surface of the positivity domain. When $\omega t$ is $0$, the positivity domain is just the whole unit sphere.}
\label{fig3}
\end{figure}

The positivity domain for each $t$ is easily found from the map of mean values
\begin{eqnarray}
  \label{eq:evol1a}
  \langle  \psig1 \rangle '& = & \apsig{1} \cos \omega t +a_1 \sin \omega t \nonumber \\
  \langle \psig2  \rangle '& = & \apsig{2} \cos \omega t + a_2 \sin \omega t \nonumber \\
\langle \psig3 \rangle' & = & \apsig{3}.
\end{eqnarray}
Regardless of whether $\avecsig$ is compatible, the density matrix for $\avecsig$, described by equation (\ref{eq:density1}), is mapped to a positive matrix, which is the density matrix for $\avecsig '$ described by the first half of equation (\ref{eq:density2}), if
\begin{equation}
  \label{eq:positd1}
  (\apsig{1} ')^2 + (\apsig{2} ')^2 + (\apsig{3}')^2 \leq 1
\end{equation}
which means $\avecsig '$ is on or inside the unit sphere described by
\begin{equation}
  \label{eq:positd2}
  \apsig{1} ' = \sin \theta \cos \varphi \, , \quad \apsig{2} ' = \sin \theta \sin \varphi \, , \quad \apsig{3} ' = \cos \theta
\end{equation}
with $\theta, \varphi$ varying over all directions. Then $\avecsig$ is on or inside the surface described by
\begin{eqnarray}
  \label{eq:positd3}
  \langle  \psig1 \rangle & = & -a_1 \tan \omega t + \frac{\sin \theta \cos \varphi}{\cos \omega t}  \nonumber \\
  \langle \psig2  \rangle & = & -a_2 \tan \omega t + \frac{\sin \theta \sin \varphi}{\cos \omega t} \nonumber \\
\langle\psig3  \rangle & = & \cos \theta
\end{eqnarray}
which is obtained from the unit sphere by moving the center distances $-a_1 \tan \omega t$ and $- a_2 \tan \omega t$ in the $x$ and $y$ directions and stretching the $x$ and the $y$ dimensions by a factor of $1/\cos \omega t$. The positivity domain is the intersection of this surface and its interior with the unit sphere and its interior, since $\avecsig$ must also be on or inside the unit sphere. The positivity domain for different values of $\omega t$ is shown in Figure \ref{fig3}.
When $\omega  t$ is $\pi/2$, the restriction (\ref{eq:positd1}) is just that
\begin{equation}
  \label{eq:positd4}
  \apsig{3} ^2 \leq 1 - (a_1) ^2 -(a_2)^2.
\end{equation}
Then the positivity domain is the part of the unit sphere where $\apsig{3}^2$ is within this bound. If $a_1$ and $a_2$ are not both zero, and $t$ is not zero, the positivity domain does not include the north pole point that corresponds to the matrix $P$ of equation (\ref{eq:op1}). 

If $a_1$ and $a_2$ are both zero, the positivity domain is the entire interior and surface of the unit sphere. Then the map takes every density matrix to a density matrix and every positive matrix to a positive matrix. In fact the map is completely positive for all $t$. The two eigenvalues of ${\cal{B}}$ that are generally negative, $\lambda_3$ and $\lambda_4$, are zero, so $C(3)$ and $C(4)$ are zero. That leaves two positive eigenvalues
\begin{equation}
  \label{eq:positd5}
  \lambda_1 = 1 + \cos \omega t \, , \qquad \lambda_2 = 1- \cos \omega t
\end{equation}
and just
\begin{equation}
  \label{eq:positd5a}
  C(1) = \sqrt{\frac{1+ \cos \omega t}{2}} \, , \qquad C(2) = \sqrt{\frac{1- \cos \omega t}{2}} \psig3.
\end{equation}

Consider three sets: the intersection of all the positivity domains for different $t$, the compatibility domain, and the set of $\avecsig$ described by the inequality (\ref{eq:domain26}). We know these sets are nested; the intersection of the positivity domains contains the compatibility domain because every positivity domain contains the compatibility domain, and we showed that the compatibility domain contains the set of $\avecsig$ described by (\ref{eq:domain26}). Now we will show that these three sets are the same; we will show that every point on the boundary of the set of $\avecsig$ described by (\ref{eq:domain26}) is also on the boundary of a positivity domain for some $t$.

In terms of the $\apsig{+}$, $\apsig{-}$ used to describe the compatibility domain, the equations (\ref{eq:positd3}) for $\avecsig$ on the boundary of the positivity domain for time $t$ are
\begin{eqnarray}
	\apsig{+} & = & -\frac{\sin \theta}{\cos \omega t} \sin (\varphi - \alpha) \nonumber \\
	\apsig{-} & = & \apsigxi{+}{1} \tan \omega t - \frac{\sin \theta}{\cos \omega t} \cos (\varphi - \alpha) \nonumber \\
	\apsig{3} & = & \cos \theta
\end{eqnarray}
with 
\begin{equation}
	a_1 = \apsigxi{+}{1} \cos \alpha \quad, \quad a_2 = \apsigxi{+}{1} \sin \alpha .
\end{equation}
If
\begin{equation}
	\label{eq:domain34}
	\sin \omega t = \frac{\apsigxi{+}{1} \cos (\varphi - \alpha)}{\sin \theta} = \frac{\apsigxi{+}{1} \cos (\varphi - \alpha)}{\sqrt{1 - \apsig{3}^2}}
\end{equation}
then
\begin{eqnarray}
	\label{eq:domain35}
	\apsig{+} & = & - \sin \theta \sin \beta = - \sqrt{1 - \apsig{3}^2} \sin \beta \nonumber \\
	\apsig{-} & = & - \sqrt{\sin^2 \theta - \apsigxi{+}{1}^2} \cos \beta = - \sqrt{1-\apsig{3}^2 - \apsigxi{+}{1}^2} \cos \beta
\end{eqnarray}
where
\begin{eqnarray}
	\label{eq:domain36}
	\sin \beta & = & \frac{\sin ( \varphi - \alpha)}{\cos \omega t} \nonumber \\
	\cos \beta & = & \frac{\sqrt{\sin^2 \theta - \apsigxi{+}{1}^2} \cos (\varphi - \alpha)}{\sin \theta \cos \omega t} \nonumber \\
	\tan \beta & = & \frac{\sin \theta}{\sqrt{\sin^2 \theta - \apsigxi{+}{1}^2}}\tan (\varphi - \alpha).
\end{eqnarray}
You can check that the sum of the squares of the formulas for $\sin \beta$ and $\cos \beta $ is $1$, so the designations $\sin \beta$ and $\cos \beta$ are allowed. Each $\avecsig$ described by these equations is on the boundary of a positivity domain. Equations (\ref{eq:domain35}) also describe the ellipses of (\ref{eq:domain28}) that are the contours of the boundary of the set of $\avecsig$ described by the inequality (\ref{eq:domain26}). From equations (\ref{eq:domain36}) we see that all values of $\beta$ from $0$ to $2 \pi$ are included as $\varphi-\alpha$ varies from $0$ to $2 \pi$, so the whole of each ellipse is included. The bound (\ref{eq:domain30}) on $\apsig{3}^2$ ensures that (\ref{eq:domain34}) does not ask $|\sin \omega t|$ to be larger than $1$ for any $\avecsig$ that satisfies (\ref{eq:domain26}), so all the ellipses of (\ref{eq:domain28}) are included. Every point on the boundary of the set of $\avecsig$ described by (\ref{eq:domain26}) is on the boundary of a positivity domain. This completes our proof that the compatibility domain and the intersection of the positivity domains both are the set of $\avecsig$ described by the inequality (\ref{eq:domain26}).

\section{General Forms}\label{general}

Consider a quantum system described by $N \times N$ matrices. The $N \times N$ Hermitian matrices  form a real linear space of $N^2$ dimensions with inner product
\begin{equation}
  \label{eq:general1}
  (A \, , \; B) = {\mbox{Tr}}[A^{\dagger}B] = \sum_{j, k =1}^N A_{kj}^* B_{kj} .
\end{equation}
Taking $N^2$ linearly independent Hermitian matrices that include the unit matrix $1$, orthogonalizing them with a Gram-Schmidt process using the inner product (\ref{eq:general1}), starting with the unit matrix, and multiplying by positive numbers for normalization, yields $N^2$ Hermitian matrices $F_{\mu 0} $ for $\mu = 0, 1, \ldots N^2-1$ such that $F_{00}$ is $1$ and
\begin{equation}
  \label{eq:general2}
  {\mbox{Tr}}[F_{\mu 0 } F_{\nu 0} ] = N \delta_{\mu \nu}.
\end{equation}
Every $N \times N$ matrix is a linear combination of the matrices $F_{\mu 0}$. 

A state of this quantum system is described by a density matrix
\begin{equation}
  \label{eq:general3}
  \rho = \frac{1}{N} \left( 1 + \sum_{\nu =1}^{N^2-1} f_{\nu} F_{\nu 0} \right).
\end{equation}
Equations (\ref{eq:general2}) imply that
\begin{equation}
  \label{eq:general4}
  \langle F_{\mu0} \rangle = {\mbox{Tr}}[F_{\mu 0} \rho] = f_{\mu}
\end{equation}
for $\mu = 1, 2, \ldots N^2 -1$, so
\begin{equation}
  \label{eq:general6}
  \rho = \frac{1}{N} \left( 1 + \sum_{\alpha =1}^{N^2-1} \langle F_{\alpha 0} \rangle  F_{\alpha 0} \right).
\end{equation}
Knowing $\rho$ is equivalent to knowing the $N^2 -1$ mean values $\langle F_{\mu 0} \rangle$ for $\mu = 1, 2, \ldots N^2 -1$. The state is described either by the density matrix or by these mean values. We can see how the state changes in time by learning how these mean values change in time. 

Suppose this first system is entangled with and interacting with a second system described by $M \times M$ matrices. Let $F_{0\mu}$ for $\mu =  0,1, \ldots M^2 -1$ be Hermitian $M \times M$ matrices such that $F_{00}$ is $1$ and
\begin{equation}
  \label{eq:general7}
  {\mbox{Tr}}[F_{0\mu} F_{0 \nu}] = M \delta_{\mu \nu}.
\end{equation}
The combined system is described by $NM \times NM$ matrices. Every $NM \times NM$ matrix is a linear combination of the matrices $F_{\mu 0} \otimes F_{0 \nu}$ which are Hermitian and linearly independent. We use notation that identifies $F_{\mu 0}$ with $F_{\mu 0} \otimes 1$ and $F_{0 \nu}$ with $1 \otimes F_{0 \nu}$ and let 
\begin{equation}
  \label{eq:general8}
  F_{\mu \nu} = F_{\mu 0} \otimes F_{0 \nu}.
\end{equation}
For these $NM \times NM$ matrices
\begin{equation}
  \label{eq:general9}
  {\mbox{Tr}}[F_{\mu \nu} F_{\alpha \beta}] = NM \; \delta_{\mu \alpha} \delta_{\nu \beta}.
\end{equation}

In the Heisenberg picture, evolution produced by a Hamiltonian $H$ for the combined system changes each matrix $F_{\mu \nu}$ to a matrix
\begin{equation}
  \label{eq:general10}
  e^{iHt}F_{\mu \nu} e^{-iHt} = \sum_{\alpha=0}^{N^2 -1} \sum_{\beta =0}^{M^2 -1} t_{\mu \nu; \alpha \beta} F_{\alpha \beta}  
\end{equation}
with real $t_{\mu \nu; \alpha \beta}$. Since
\begin{equation}
  \label{eq:general11}
  {\mbox{Tr}} \left[  e^{iHt}F_{\mu \nu} e^{-iHt} e^{iHt}F_{\alpha \beta} e^{-iHt} \right] =  {\mbox{Tr}}[F_{\mu \nu} F_{\alpha \beta}],
\end{equation}
the $t_{\mu \nu; \alpha \beta}$ form an orthogonal matrix, so $t^{-1}_{ \alpha \beta; \mu \nu}$ is  $t_{\mu \nu; \alpha \beta}$ and
\begin{equation}
  \label{eq:general12}
  e^{-iHt}F_{\alpha \beta} e^{iHt}=   \sum_{\mu=0}^{N^2 -1} \sum_{\nu =0}^{M^2 -1} t_{\mu \nu; \alpha \beta} F_{\mu \nu}. 
\end{equation}
Since $F_{00}$ is $1$,
\begin{equation}
  \label{eq:general13}
  t_{00; \alpha \beta} = \delta_{0 \alpha} \delta_{0 \beta} \, , \qquad t_{\mu \nu ; 00} = \delta_{\mu 0} \delta_{\nu 0} .
\end{equation}
Forming an orthogonal matrix is not the only property the $t_{\mu \nu; \alpha \beta}$ need to have. They must also yield
\begin{equation}
  \label{eq:general14}
    e^{iHt}F_{\mu \nu} e^{-iHt}=  e^{iHt}F_{\mu 0} e^{-iHt}  e^{iHt}F_{0 \nu} e^{-iHt}
\end{equation}
and the same with $t$ changed to $-t$.

The mean values $\langle F_{\mu 0} \rangle$ for $\mu = 1,2, \ldots N^2-1$ that describe the state of the first system at time zero are changed to the mean values 
\begin{equation}
  \label{eq:general15}
  \langle F_{\mu 0} \rangle' = \langle e^{iHt}F_{\mu 0} e^{-iHt} \rangle = d_{\mu} + \sum_{\alpha=1}^{N^2 -1} t_{\mu 0 ; \alpha 0}   \langle F_{\alpha 0} \rangle
\end{equation}
that describe the state of the first system at time $t$, with
\begin{equation}
  \label{eq:general16}
  d_{\mu}= \sum_{\alpha=0}^{N^2 -1} \sum_{\beta=1}^{M^2 -1} t_{\mu 0 ; \alpha \beta}   \langle F_{\alpha \beta} \rangle.
\end{equation}
Mean values $\langle F_{\alpha  0} \rangle$ that describe the state of the first system are in equation (\ref{eq:general15}) but not in (\ref{eq:general16}). We consider the $d_{\mu}$, as well as the $t_{\mu 0 ; \alpha 0}$ to be parameters that describe the effect on the first system of the dynamics of the combined system that drives the evolution of the first system, not part of the description of the initial state of the first system. 

The density matrix $\rho$ of equation (\ref{eq:general6}) that describes the state of the first system at time zero is changed to the density matrix 
\begin{equation}
  \label{eq:general17}
  \rho' = \frac{1}{N} \left( 1 + \sum_{\mu = 1}^{N^2 -1}\langle F_{\mu 0} \rangle' F_{\mu 0} \right)
\end{equation}
that describes the state at time $t$. Equations (\ref{eq:general15}) imply it is
\begin{equation}
  \label{eq:general18}
  \rho' = \frac{1}{N} \left( 1 + \sum_{\mu =1}^{N^2 -1} d_{\mu} F_{\mu 0} + \sum_{\alpha=1}^{N^2-1} \langle F_{\alpha 0} \rangle \sum_{\mu =1}^{N^2 -1}  t_{\mu 0; \alpha 0}F_{\mu 0} \right).  
\end{equation}

Equation (\ref{eq:general18}) for $\rho'$ can be obtained another way. In the Schr\"{o}dinger picture the density matrix
\begin{equation}
  \label{eq:general19}
  \Pi = \frac{1}{NM} \left( 1 + \sum_{\alpha=1}^{N^2 -1} \langle F_{\alpha 0} \rangle  F_{\alpha 0} + \sum_{\alpha=0}^{N^2 -1}\sum_{\beta=1}^{M ^2 -1}\langle F_{\alpha \beta} \rangle  F_{\alpha \beta} \right)
\end{equation}
that represents the state of the combined system at time zero is changed at time $t$ to
\begin{eqnarray}
\label{eq:general20}
e^{-iHt} \Pi e^{iHt} &=& \frac{1}{NM} \left( 1 +  \sum_{\alpha=1}^{N^2 -1} \langle F_{\alpha 0} \rangle \sum_{\mu =1}^{N^2 -1} t_{\mu 0; \alpha 0} F_{\mu 0} \right. \nonumber \\
&& + \sum_{\alpha=1}^{N^2 -1} \langle F_{\alpha 0} \rangle \sum_{\mu =0}^{N^2 -1}\sum_{\nu =1}^{M^2 -1} t_{\mu \nu; \alpha 0} F_{\mu \nu} \nonumber \\
&& +  \sum_{\alpha=0}^{N^2 -1} \sum_{\beta=1}^{M^2 -1}\langle F_{\alpha \beta} \rangle  \sum_{\mu =1}^{N^2 -1} t_{\mu 0; \alpha \beta} F_{\mu 0} \nonumber \\
&& +  \left. \sum_{\alpha=0}^{N^2 -1} \sum_{\beta=1}^{M^2 -1}\langle F_{\alpha \beta} \rangle \sum_{\mu =0}^{N^2 -1}\sum_{\nu =1}^{M^2 -1} t_{\mu \nu; \alpha \beta} F_{\mu \nu} \right)
\end{eqnarray}
according to equations (\ref{eq:general12}). Taking the partial trace of this over the states of the second system eliminates the $F_{\mu \nu}$ for $\nu$ not zero and gives equation (\ref{eq:general18}) for the density matrix of the first system at time $t$ with equations (\ref{eq:general16}) for the $d_{\mu}$. Since this involves working with the larger system longer, it does not appear to be the easier way to actually do a calculation.

The map from density matrices (\ref{eq:general6}) at time zero to density matrices (\ref{eq:general18}) at time $t$ holds for all the varying mean values $\langle F_{\alpha 0} \rangle$ that are compatible with fixed mean values  $\langle F_{\alpha \beta} \rangle$ in the $d_{\mu}$ in describing a possible initial state for the combined system. We will refer to them as compatible  $\langle F_{\alpha 0} \rangle$. Almost all initial states of the combined system allow the compatible  $\langle F_{\alpha 0} \rangle$ to vary as $N^2 -1$ independent variables. We will consider only those initial states. 

The map of density matrices extends to a linear map of all $N \times N$ matrices to $N \times N$ matrices defined by 
\begin{equation}
  \label{eq:mapgen}
  1' = 1 + \sum_{\mu =1}^{N^2 -1} d_{\mu} F_{\mu 0} \, , \qquad F_{\alpha 0} ' = \sum_{\mu =1}^{N^2 -1} t_{\mu 0 ; \alpha 0} F_{\mu 0}.
\end{equation}
It takes the density matrix (\ref{eq:general6}) to the density matrix (\ref{eq:general18}) for each of the varying compatible $\langle F_{\alpha 0} \rangle$. It takes every Hermitian matrix to a Hermitian matrix.

The latter property alone is the foundation for basic forms of the map. This statement is independent of our other considerations.

\noindent
{\bf Lemma:} If a linear map $Q \rightarrow Q'$ of $N \times N$ matrices to $N \times N$ matrices maps every Hermitian matrix to a Hermitian matrix, then in the description of the map by
\begin{equation}
  \label{eq:lemma1}
  Q'_{rs} = \sum_{j,k=1}^N {\cal{B}}_{rj; sk}Q_{jk} 
\end{equation}
the $N^2 \times N^2$ matrix ${\cal{B}}$ is uniquely determined by the map and is Hermitian,
\begin{equation}
  \label{eq:lemma2}
  {\cal{B}}^*_{rj; sk} = {\cal{B}}_{sk; rj},
\end{equation}
and there are $N \times N$ matrices $C(n)$ for $n = 1, \ldots N^2$ such that
\begin{equation}
  \label{eq:lemma3}
  Q' = \sum_{n=1}^p C(n) Q C(n)^{\dagger} - \sum_{n=p+1}^{N^2} C(n) Q C(n)^{\dagger}
\end{equation}
for all $Q$, and
\begin{equation}
  \label{eq:lemma4}
  {\mbox{Tr}}[C(m)^{\dagger} C(n)] = 0
\end{equation}
for $m \neq n$, for $m,n = 1, \ldots N^2$.

\noindent
{\bf Proof:} Let $E_{jk}$ be the $N \times N$ matrices defined by 
\begin{equation}
  \label{eq:lemma5}
  [E_{jk}]_{lm}=\delta_{lj} \delta_{mk}.
\end{equation}
Clearly $E^{\dagger}_{jk} = E_{kj}$. If the map takes every Hermitian matrix to a Hermitian matrix, then $({\mbox{Re}}[E_{jk}])'$ and $({\mbox{Im}}[E_{jk}])'$ are Hermitian and
\begin{equation}
  \label{eq:lemma6}
 \{ (E_{jk})'\}^{\dagger}=\{ ({\mbox{Re}}[E_{jk}])' + i ({\mbox{Im}}[E_{jk}])' \}^{\dagger} = ({\mbox{Re}}[E_{jk}])' - i ({\mbox{Im}}[E_{jk}])' =(E^{\dagger}_{jk})'.
\end{equation}
Equations (\ref{eq:lemma1}) and (\ref{eq:lemma5}) give
\begin{equation}
  \label{eq:lemma7}
  [E_{jk}']_{rs} = \sum_{l,m} {\cal{B}}_{rl;sm} \delta_{lj} \delta_{mk} = {\cal{B}}_{rj;sk} 
\end{equation}
which shows that the map determines a unique ${\cal{B}}$, and with
\begin{equation}
  \label{eq:lemma8}
  (E_{jk}')^{\dagger} = (E^{\dagger}_{jk})' = E_{kj}'
\end{equation}
implies that ${\cal{B}}^*_{sj;rk} = {\cal{B}}_{rk;sj}$ which is the same as equation (\ref{eq:lemma2}).

Since ${\cal{B}}$ is Hermitian, it has a spectral decomposition
\begin{equation}
  \label{eq:lemma9}
  {\cal{B}} = \sum_{n=1}^{N^2} \lambda_n | n \rangle \langle n | 
\end{equation}
where the $| n \rangle$ are orthonormal eigenvectors of ${\cal{B}}$ and the $\lambda_n$ are eigenvalues. The $\lambda_n$ are real, but they are not necessarily all different, non-zero, or non-negative. We label them so that
\begin{equation}
  \label{eq:lemma10}
  \lambda_n \geq 0 \quad {\mbox{for}} \quad n = 1, \ldots p \qquad ; \qquad \lambda_n \leq 0 \quad {\mbox{for}} \quad n = p+1, \ldots N^2.
\end{equation}
Then
\begin{equation}
  \label{eq:lemma11}
  {\cal{B}}_{rj;sk} = \sum_{n=1}^{p} \sqrt{|\lambda_n|}\langle r \, j | n \rangle \langle n | s \, k \rangle \sqrt{|\lambda_n|} - \sum_{n=p+1}^{N^2} \sqrt{|\lambda_n|}\langle r \, j | n \rangle \langle n | s \, k \rangle \sqrt{|\lambda_n|}.
\end{equation}
Let
\begin{equation}
  \label{eq:lemma12}
  C(n)_{rj} =  \sqrt{|\lambda_n|}\langle r \, j | n \rangle.
\end{equation}
Then equation (\ref{eq:lemma1}) is
\begin{equation}
  \label{eq:lemma13}
  Q'_{rs} = \sum_{n=1}^{p} \sum_{jk} C(n)_{rj} Q_{jk} C(n)_{sk}^* -\sum_{n=p+1}^{N^2} \sum_{jk} C(n)_{rj} Q_{jk} C(n)_{sk}^*
\end{equation}
so the map is described by equation (\ref{eq:lemma3}), and
\begin{equation}
  \label{eq:lemma14}
  {\mbox{Tr}}[C(m)^{\dagger} C(n)] = \sum_{rj} C(m)_{rj}^*C(n)_{rj} = \sum_{rj} \sqrt{|\lambda_m|} \langle m | r\, j \rangle \langle r \, j| n \rangle \sqrt{|\lambda_n|} = | \lambda_n| \langle  m | n \rangle
\end{equation}
which is zero for $m \neq n$ in accord with equation (\ref{eq:lemma4}).

This completes the proof of the Lemma.

The maps we are considering, those described by equations (\ref{eq:mapgen}), have the additional property that
\begin{equation}
  \label{eq:lemma15}
  {\mbox{Tr}} \; Q' = {\mbox{Tr}} \; Q
\end{equation}
for every $Q$. This implies that
\begin{equation}
  \label{eq:lemma16}
  \sum_{n=1}^{p}C(n)^{\dagger}C(n) -  \sum_{n=p+1}^{N^2}C(n)^{\dagger}C(n)=1
\end{equation}
because 
\begin{equation}
  \label{eq:lemma17}
   {\mbox{Tr}}\; Q  = {\mbox{Tr}}\; Q'= {\mbox{Tr}} \left[ \left(  \sum_{n=1}^{p}C(n)^{\dagger}C(n) -  \sum_{n=p+1}^{N^2}C(n)^{\dagger}C(n) \right)Q \right]
\end{equation}
implies that in the linear space of $N \times N$ matrices with the inner product defined by the trace as in (\ref{eq:general1}), the difference between the two sides of equation (\ref{eq:lemma16}) has zero inner product with every matrix $Q$ and therefore must be zero. From equations (\ref{eq:lemma5}) and (\ref{eq:lemma7}) we see also that the trace-preserving property described by (\ref{eq:lemma15}) implies that
\begin{equation}
  \label{eq:lemma18}
  \sum_r {\cal{B}}_{rj; rk} = {\mbox{Tr}}[E_{jk}'] = {\mbox{Tr}} [E_{jk}] = \delta_{jk}.
\end{equation}
Conversely, either equation (\ref{eq:lemma16}) or (\ref{eq:lemma18}) implies that ${\mbox{Tr}}\; Q'$ equals ${\mbox{Tr}}\; Q$ for every matrix $Q$. From equation (\ref{eq:lemma18}) we see in particular that ${\mbox{Tr}}\; {\cal{B}}$ is $N$. 

\section{Discussion}

In the light of understanding gained here, it is easy to see the errors in arguments that a map describing evolution of an open quantum system {\em has to be completely positive}. One argument uses the fact that a map for a system $A$ is completely positive if and only if it is the contraction to $A$ of unitary evolution of a larger system $AB$ in which $A$ is combined with another system $B$ and the density matrix for the initial state of $AB$ is a product of density matrices for $A$ and $B$. That is clearly not necessary.

Another argument uses the fact that a map for a system $A$ is completely positive if and only if the product of that map with the identity map for another system $C$ yields a map for the combined system $AC$ that takes every positive matrix for $AC$ to a positive matrix. The argument says this is the way to satisfy the physically reasonable requirement that the description of the evolution of $A$ must allow $A$ to be accompanied by another system $C$ that could be entangled with $A$ but does not respond to the dynamics that drives the evolution of $A$. If the map for $A$ is a contraction to $A$ of either unitary evolution or a completely positive map for a larger system $AB$ in which $A$ is combined with another system $B$, then the evolution of $B$ is generally not described by the identity map, so $C$ is not $B$. The accompanying system $C$ must be a third system. The physically reasonable requirement can be satisfied very simply for the kind of maps we have considered. If the map for $AB$ is completely positive, its product with the identity map for $C$ yields a map for the combined system $ABC$ that takes every positive matrix for $ABC$ to a positive matrix. 

Mathematically, a map of states for a subsystem $A$ can be constructed from (1) a map that takes density matrices for $A$ to density matrices for the entire system $AB$ at time zero, followed by (2) unitary Hamiltonian evolution from time zero to time $t$ for $AB$, and finally (3) the trace over the states of $B$ that yields the density matrix for $A$ at time $t$. The broad class of maps obtained this way is known to include maps that are not completely positive and in fact maps that do not take every positive matrix to a positive matrix. That all depends on the first step, the map that assigns density matrices $\rho_{AB}$ to density matrices $\rho_A$ at time zero. Pechukas \cite{pechukas94} has shown that if $A$ is a qubit, the only linear assignment of density matrices $\rho_{AB}$ that applies to all density matrices $\rho_A$, and gives back unchanged $\rho_A$ in the trace over $B$ at time zero, is
\begin{equation}
  \label{eq:disc1}
  \rho_A \rightarrow \rho_A \otimes \rho_B
\end{equation}
with $\rho_B$ fixed. We prove this for any quantum system in an Appendix. Pechukas concludes that in general, when  product assignments (\ref{eq:disc1}) do not apply, maps have to act on limited domains. This does not depend on the unitary evolution of $AB$ from time zero to time $t$. When a product assignment (\ref{eq:disc1}) is the first step, the map made in three steps is completely positive; if a map made this way is not completely positive, its domain must be limited. There has been debate whether any except the completely positive maps can describe physical evolution \cite{alicki95,pechukas95}.
 
      Which do describe physical evolution? What is needed for one of these maps to describe evolution of states of $A$ caused by dynamics of $AB$? If the map is meant to apply to a set of $\rho_A$ that all evolve in time as a result of the same cause, the $\rho_{AB}$ assigned to these $\rho_A$ should not differ in ways that would change the cause of evolution of the $\rho_A$. If they did, we would say that different $\rho_A$ are being handled differently and that their evolution should be described by different maps. Pechkas \cite{pechukas94} considers the case where $A$ and $B$ are qubits and a product $\rho_{AB}$ is assigned, as in (\ref{eq:disc1}), to each of four selected $\rho_A$, with a different $\rho_B$ for each of the four $\rho_A$. This yields a map that takes every mixture of the four $\rho_A$ to a density matrix. Pechukas observes that the large set of maps obtained this way must include many that are not completely positive and many that take density matrices outside the set of mixtures to matrices that are not positive. However, the $\rho_{AB}$ assigned to each different mixture generally gives a different density matrix for $B$ in the trace over the states of $A$. Each different state of $A$ is coupled with a different state of $B$. Does this mean it is handled differently? If a map is meant to describe evolution that has a definite physical cause, does Pechukas  have a single map that acts on a set of states; or a set of maps, each acting on a single state?

      In the compatibility domain that we describe, the evolution of all the states is clearly the result of the same cause. It can be described by a single map that has physical meaning. Working with mean values helps make this clear. We do not need a complete description of the state of $AB$ at time zero. It does not need to stand alone, independent of the unitary evolution, and accommodate any unitary evolution. The compatibility domain depends on the unitary evolution. In our example, the compatibility domain depends on the mean values that are the parameters $a_1$ and $a_2$. That these mean values are the relevant parameters depends on our choice of Hamiltonian. The compatibility domain is unlimited when $a_1$ and $a_2$ are zero. Then the map is completely positive, but that does not require an initial state described by a density matrix that is a product. 

\appendix*
\section{Generalization of Pechukas' result}

\noindent {\bf Theorem.} If a linear map applies to all density matrices $\rho_A$ for a subsystem $A$ and assigns each $\rho_A$ a density matrix $\rho_{AB}(\rho_A)$ for the combined system $AB$ so that 
\begin{equation}
  \label{eq:app1}
  {\mbox Tr}_B [ \rho_{AB}(\rho_A)] = \rho_A ,
\end{equation}
then, for every $\rho_A$,
\begin{equation}
  \label{eq:app2}
   \rho_{AB}(\rho_A)= \rho_A \otimes \rho_B
\end{equation}
with $\rho_B$ a density matrix for the subsystem $B$ that is the {\em same} for all $\rho_A$.

\noindent {\bf Proof.} The first step, which Pechukas \cite{pechukas94} did, is to show that every pure-state density matrix $\rho_A$ is assigned a product density matrix, as in (\ref{eq:app2}), with $\rho_B$ possibly different for different $\rho_A$. For completeness we include a slightly different presentation of this step. If $\rho_A$ represents a pure state, there is an orthonormal basis of state vectors $| \psi_j \rangle$ for $A$, with $j=1,2, \ldots$, such that $\rho_A$ is $|\psi_1 \rangle \langle \psi_1|$. We combine these with orthonormal state vectors $| \phi_k \rangle$ for $B$ to make an orthonormal basis of product vectors $|\psi_j \phi_k \rangle$ for $AB$. Since $\rho_{AB}(\rho_A)$ is positive, each $\langle \psi_j \phi_k | \rho_{AB}(\rho_A)| \psi_j \phi_k \rangle$ is non-negative and, from (\ref{eq:app1}), if $j$ is not $1$,
\begin{equation}
  \label{eq:app3}
  \langle \psi_j \phi_k | \rho_{AB}(\rho_A)| \psi_j \phi_k \rangle \leq \langle \psi_j | {\mbox {Tr}}_B[\rho_{AB}(\rho_A)]| \psi_j  \rangle = \langle \psi_j | \psi_1 \rangle \langle \psi_1 | \psi_j \rangle = 0. 
\end{equation}
Since $\rho_{AB}(\rho_A)$ is positive, it is the square of a Hermitian operator. Thus we see that  $\rho_{AB}(\rho_A) |\psi_j \phi_k \rangle$ is zero if $j$ is not $1$ and 
\begin{equation}
  \label{eq:app4}
    \langle \psi_j \phi_r | \rho_{AB}(\rho_A)| \psi_k \phi_s \rangle = \delta_{j1} \delta_{k1}   \langle \psi_1 \phi_r | \rho_{AB}(\rho_A)| \psi_1 \phi_s \rangle.
\end{equation}
Let
\begin{equation}
  \label{eq:app5}
  \rho_B(\rho_A) = {\mbox Tr}_A [ \rho_{AB}(\rho_A)].
\end{equation}
Then
\begin{equation}
  \label{eq:app6}
   \rho_B(\rho_A) = \langle \psi_1|\rho_{AB}(\rho_A)| \psi_1 \rangle.
\end{equation}
and 
\begin{equation}
  \label{eq:app7}
  \rho_{AB}(\rho_A) = | \psi_1 \rangle \langle \psi_1 | \otimes \rho_B(\rho_A).
\end{equation}
That completes the first step of the proof. 

The second step, which completes the proof of the theorem, is to show that $\rho_B$ is the same for all pure-state density matrices $\rho_A$. Pechukas \cite{pechukas94} did this for the case where $A$ is a qubit. We show that the proof an be easily extended to any quantum system. Suppose $| \psi_1 \rangle$ and $|\psi_2 \rangle$ are orthonormal state vectors for $A$. Let
\begin{eqnarray}
  \label{eq:app8}
  | \psi_3 \rangle & =& \frac{1}{\sqrt{2}}| \psi_ 1\rangle + \frac{i}{\sqrt{2}}e^{i \beta} | \psi_2 \rangle \nonumber \\
 | \psi_4 \rangle & =& \frac{1}{\sqrt{2}}| \psi_ 1\rangle - \frac{i}{\sqrt{2}}e^{i \beta} | \psi_2 \rangle \nonumber \\
 | \psi_5 \rangle & =& \cos \alpha \, | \psi_ 1\rangle + \sin \alpha \, e^{i \beta} | \psi_2 \rangle \nonumber \\
| \psi_6 \rangle & =& \sin \alpha \, | \psi_ 1\rangle - \cos \alpha \, e^{i \beta} | \psi_2 \rangle.
\end{eqnarray}
Then $| \psi_3 \rangle$ and $|\psi_4 \rangle$ are orthogonal, $| \psi_5 \rangle$ and $|\psi_6 \rangle$ are orthogonal, and $|\langle \psi_1 | \psi_3 \rangle |^2 $, $|\langle \psi_1 | \psi_4 \rangle |^2 $, $|\langle \psi_2 | \psi_3 \rangle |^2 $, $|\langle \psi_2 | \psi_4 \rangle |^2 $, $|\langle \psi_3 | \psi_5 \rangle |^2 $, $|\langle \psi_3 | \psi_6 \rangle |^2 $, $|\langle \psi_4 | \psi_5 \rangle |^2 $, and $|\langle \psi_4 | \psi_6 \rangle |^2 $ are all $1/2$. The length of each vector $|\psi_k \rangle$ is $1$, so $|\psi_k \rangle \langle \psi_k |$ is a pure-state density matrix for $A$. The map assigns it a product density matrix
\begin{equation}
  \label{eq:app9}
  \rho_{AB} (|\psi_k \rangle \langle \psi_k |) = |\psi_k \rangle \langle \psi_k | \otimes \rho_B(k)
\end{equation}
as in (\ref{eq:app7}) with $\rho_B(k)$ short notation for $\rho_B(|\psi_k \rangle \langle \psi_k |)$.

Since the map is linear, it follows from
\begin{equation}
  \label{eq:app10}
  \frac{1}{2}(|\psi_1 \rangle \langle \psi_1 | + |\psi_2 \rangle \langle \psi_2 | ) = \frac{1}{2}(|\psi_3 \rangle \langle \psi_3 | + |\psi_4 \rangle \langle \psi_4 | ) 
\end{equation}
that
\begin{equation}
  \label{eq:app11}
\frac{1}{2}(|\psi_1 \rangle \langle \psi_1 |\, \rho_B(1)  + |\psi_2 \rangle \langle \psi_2 | \, \rho_B (2) ) = \frac{1}{2}(|\psi_3 \rangle \langle \psi_3 | \, \rho_B(3) + |\psi_4 \rangle \langle \psi_4 |\, \rho_B(4) ).
\end{equation}
Taking partial mean values $\langle \psi_1| \ldots | \psi_1 \rangle$, $\langle \psi_2| \ldots | \psi_2 \rangle$, $\langle \psi_3| \ldots | \psi_3 \rangle$ of this last equation (\ref{eq:app11}) yields three equations that imply $\rho_B(1)$, $\rho_B(2)$, $\rho_B(3)$ and $\rho_B(4)$ all are the same. Doing everything starting from (\ref{eq:app10}) again with $1, 2, 3,4$ changed to $3,4,5,6$ shows that $\rho_B(3)$, $\rho_B(4)$, $\rho_B(5)$ and $\rho_B(6)$ all are the same. Any state vector for $A$ is in a subspace spanned by $|\psi_1 \rangle$ and a vector $|\psi_2 \rangle$  orthogonal to $| \psi_1 \rangle$, so $|\psi_5\rangle$ with fixed $|\psi_1 \rangle$ and varying $\alpha$, $\beta$ and $|\psi_2 \rangle$ can represent any pure state for $A$. If $\rho_A$ represents a pure state, $\rho_B(\rho_A)$ is the same as $\rho_B (1)$, so (\ref{eq:app2}) holds, with the same $\rho_B$, for all pure states of $A$ and, therefore, for all mixtures as well. This completes the proof of the theorem. 
\bibliography{ncp}


\end{document}